\documentstyle[12pt]{article}
\begin{document}
\title{THE SCALED UNIVERSE II}
\author{B.G. Sidharth$^*$\\
B.M. Birla Science Centre, Hyderabad 500 063 (India)}
\date{}
\maketitle
\footnotetext{$^*$E-mail:birlasc@hd1.vsnl.net.in}
\begin{abstract}
In an earlier paper we had pointed out that Quantum Mechanical type effects are
seen at different scales in the macro universe also. In this paper we obtain
a rationale for this, which lies in the picture of bound material systems,
spanning a Compton wavelength type extent, separated by much larger and relatively
much less dense distances.
\end{abstract}
\section{Introduction}
In a previous communication\cite{r1} it was shown that the mysterious quantum
prescription of microphysics has analogues at the much larger scale of stars,
galaxies and superclusters. The common feature in all these cases is a Brownian
type fractality. We now examine this circumstance in greater detail to obtain
a rationale for this "Scaled Quantum Mechanical" behaviour.\\
At the outset it must be observed that in a sense, the universe is not a
continuum.
Thus within an atom, the nucleus occupies a very tiny fraction
of the volume, roughly $\sim 10^{-15}$. Then there is a wide gap till we
reach the orbiting electrons. Similarly there are intermolecular distances,
interplanetary, interstellar, intergalactic.... distances which provide
relatively huge gaps. This is not in the spirit of a uniform continuum.\\
We will now argue that this is because, for example the nucleons are bound
together, so also the electrons and the nucleons are bound together, the
atoms in the molecules are bound together... and so on with subsequent gaps, which leads to some
interesting scale dependent consequences, all this in the context of a
Brownian underpinning.
\section{A Rationale for Scaled Effects}
The starting point in\cite{r1} was the fact that we have the following
Random Walk type Relations:
\begin{equation}
R \approx l_1 \sqrt{N_1}\label{e1}
\end{equation}
\begin{equation}
R \approx l_2 \sqrt{N_2}\label{e2}
\end{equation}
\begin{equation}
l_2 \approx l_3 \sqrt{N_3}\label{e3}
\end{equation}
\begin{equation}
R \sim l\sqrt{N}\label{e4}
\end{equation}
where $N_1 \sim 10^6$ is the number of superclusters in the universe,
$l_1 \sim 10^{25}cms$ is a typical supercluster size $N_2 \sim 10^{11}$ is the
number of galaxies in the universe and $l_2  \sim 10^{23}cms$ is the typical size
of a galaxy, $l_3 \sim 1$ light year is a typical distance between stars and
$N_3 \sim 10^{11}$ is the number of stars in a galaxy, $R$ being the radius
of the universe $\sim 10^{28} cms, N \sim 10^{80}$ is the number of elementary
particles, typically pions in the universe and $l$ is the pion Compton
wavelength.\\
On this basis it was argued that we could introduce a "Scaled" Planck Constant
given by
\begin{equation}
h_1 \sim 10^{93}\label{e5}
\end{equation}
for super clusters;
\begin{equation}
h_2 \sim 10^{74}\label{e6}
\end{equation}
for galaxies and
\begin{equation}
h_3 \sim 10^{54}\label{e7}
\end{equation}
for stars.\\
The relation (\ref{e4}) was observed nearly
a century ago by Weyl and Eddington.
It was shown (Cf.ref.\cite{r2}) that far from being empirical this relation
can be deduced on the basis of the fluctuational creation of particles from a
background Zero Point Field or Quantum Vacuum, in a scheme which leads to a
cosmology consistent also with Dirac's large number coincidences\cite{r3}.
From this point
of view the Random Walk character of equation (\ref{e4}) is not accidental, and
this reasoning could be extended to equations (\ref{e1}),(\ref{e2}) and
(\ref{e3}), in the light of equations (\ref{e5}), (\ref{e6}) and (\ref{e7}).\\
It may be observed that in all these cases we have a length, the Compton
wavelength or its analogue which defines regions of matter separating
relatively empty spaces.\\
Infact it was also argued in\cite{r1} that these scaled "Compton wavelengths"
and scaled "Planck Constants" arise due to the well known equation of
gravitational orbits,
\begin{equation}
\frac{GM}{L} \sim v^2\label{e8}
\end{equation}
On the other hand equation (\ref{e8}) can be viewed as resulting from the
Virial Theorem\cite{r4}, where the velocity is replaced by the velocity
dispersion.\\
This velocity $v$ would be different at different scales. For example for a black hole
it would be the velocity of light, which would then give the Schwarzchild
radius from (\ref{e8}). For galaxies it is $\sim 10^7 cms$
per second\cite{r5}. It is this circumstance that produces the above scales
leading to fractality.\\
We could go one step further, because we expect that the same effect would
apply to solar type systems: The planets and other objects are bound quite
close to the sun compared to the interstellar distances. Infact we can
verify that this is so for Kuiper Belt objects which have been studied in
the recent past\cite{r6}. In this case a typical size is $\sim 5 km$, the
distances are $\sim 10^{15}cm$, masses are $\sim 10^{19}gms$ their number
is $\sim 10^{10}$ while an application of equation (\ref{e8}) shows that the
velocities are $\sim 10^5 cms$ per second which is indeed so. It can now be shown quite easily
that this defines a scaled Planck Constant
$$h_4 \sim 10^{34}$$
In other words
equation (\ref{e8}) characterizes "Black Holes" with different maximal
velocities.\\
Incidentally from (\ref{e8}) we could easily deduce that the angular momentum
$J$ is given by
\begin{equation}
J \propto M^2\label{e9}
\end{equation}
It is quite remarkable that the equation (\ref{e9}) also applies to elementary
particles and Regge trajectories\cite{r7}.\\
Indeed this is no coincidence as can be seen as follows: It was shown that
a pion can be considered to be a bound state of an electron and a positron
within the context of Quantum Mechanical Black Holes\cite{r8}. In this case we
have instead of equation (\ref{e8})
\begin{equation}
c^2 \sim \frac{e^2}{m_e r}\label{e10}
\end{equation}
where $m_e$ is the electron mass.\\
Whence we have
\begin{equation}
h \approx m_\pi cr \sim \frac{e^2 \times 10^2}{c} \sim 10^{-27}\label{e11}
\end{equation}
It is remarkable that (\ref{e11}) gives us the value of Planck's constant,
given the pion and electron masses, and at the same time we can deduce an
equation like (\ref{e9}) at the micro level also. Equation (\ref{e11}) is on the
same footing as equations (\ref{e5}), (\ref{e6}) and (\ref{e7}). In this case
it is the electromagnetic interaction which replaces the gravitational
interaction in the equation (\ref{e8}).\\
Thus the rationale for the fractal structure seen above is in bound systems
separated by relatively large and relatively empty spaces.

\end{document}